\documentclass[showpacs, showkeys, amsmath, amssymb, notitlepage, superscriptaddress, reprint, aip]{revtex4-1}

\usepackage[normalem]{ulem}
\usepackage[usenames,dvipsnames]{color}
\usepackage{upgreek}
\usepackage{dcolumn}

\usepackage{graphicx}
\usepackage{bm}
\usepackage{siunitx}
\usepackage[utf8]{inputenc}
\usepackage[T1]{fontenc}
\usepackage{mathptmx}
\usepackage{textcomp}
\usepackage{hyperref}

\begin{document}

\preprint{AIP/123-QED}
\title{Coherent superconducting qubits from a subtractive junction fabrication process}

\author{Alexander Stehli}
\email[corresponding author: ]{alexander.stehli@kit.edu}
\affiliation{Institute of Physics, Karlsruhe Institute of Technology, 76131 Karlsruhe, Germany}
\author{Jan David Brehm}
\affiliation{Institute of Physics, Karlsruhe Institute of Technology, 76131 Karlsruhe, Germany}
\author{Tim Wolz}
\affiliation{Institute of Physics, Karlsruhe Institute of Technology, 76131 Karlsruhe, Germany}
\author{Paul Baity}
\affiliation{James Watt School of Engineering, University of Glasgow, Glasgow G12 8LT, United Kingdom}
\author{Sergey Danilin}
\affiliation{James Watt School of Engineering, University of Glasgow, Glasgow G12 8LT, United Kingdom}
\author{Valentino Seferai}
\affiliation{James Watt School of Engineering, University of Glasgow, Glasgow G12 8LT, United Kingdom}
\author{Hannes Rotzinger}
\affiliation{Institute of Physics, Karlsruhe Institute of Technology, 76131 Karlsruhe, Germany}
\affiliation{Institute for Quantum Materials and Technologies, Karlsruhe Institute of Technology, 76344 Eggenstein-Leopoldshafen, Germany}
\author{Alexey V. Ustinov}
\affiliation{Institute of Physics, Karlsruhe Institute of Technology, 76131 Karlsruhe, Germany}
\affiliation{National University of Science and Technology MISIS, Moscow 119049, Russia}
\affiliation{Russian Quantum Center, Skolkovo, Moscow 143025, Russia}
\author{Martin Weides}
\affiliation{James Watt School of Engineering, University of Glasgow, Glasgow G12 8LT, United Kingdom}

\date{\today}

\begin{abstract}
	Josephson tunnel junctions are the centerpiece of almost any superconducting electronic circuit, including qubits. Typically, the junctions for qubits are fabricated using shadow evaporation techniques to reduce dielectric loss contributions from the superconducting film interfaces. In recent years, however, sub-micron scale overlap junctions have started to attract attention. Compared to shadow mask techniques, neither an angle dependent deposition nor free-standing bridges or overlaps are needed, which are significant limitations for wafer-scale processing. This comes at the cost of breaking the vacuum during fabrication, but simplifies integration in multi-layered circuits, implementation of vastly different junction sizes, and enables fabrication on a larger scale in an industrially-standardized process.
	In this work, we demonstrate the feasibility of a subtractive process for fabrication of overlap junctions. In an array of test contacts, we find low aging of the average normal state resistance of only 1.6\% over 6 months. We evaluate the coherence properties of the junctions by employing them in superconducting transmon qubits. In time domain experiments, we find that both, the qubit life- and coherence time of our best device, are on average greater than $20\,\si{\micro\second}$. Finally, we discuss potential improvements to our technique. This work paves the way towards a more standardized process flow with advanced materials and growth processes, and constitutes an important step for large scale fabrication of superconducting quantum circuits.
\end{abstract}

\maketitle
\begin{figure}
	\centering
	\includegraphics{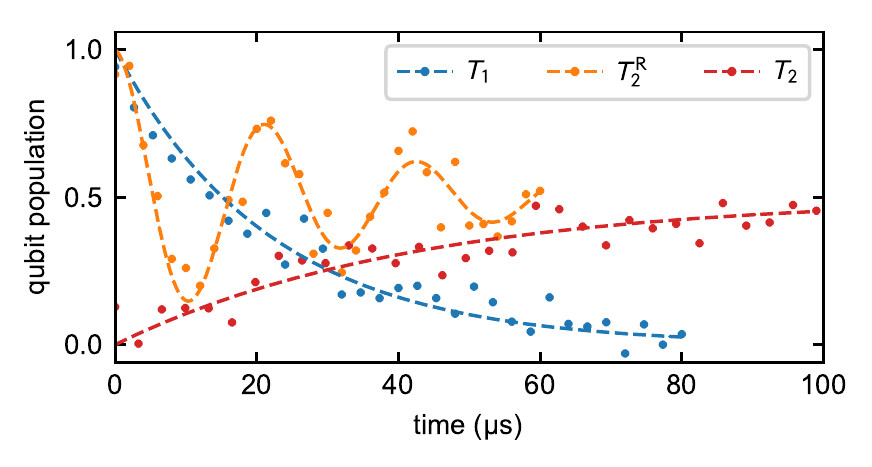}
	\caption{Qubit population during a typical decay time measurement. The dashed lines represent a fit to the qubit's lifetime ${T_1 = (22.0\pm0.9)\,\si{\micro\second}}$, and coherence times $T_2^{\rm R} = (30.1\pm3.7)\,\si{\micro\second}$ (Ramsey experiment) and $T_2 = (42.3\pm2.9)\,\si{\micro\second}$ (spin-echo experiment).}
	\label{fig:exemplary_decay_times}
\end{figure}
\begin{figure}
	\centering
	\includegraphics{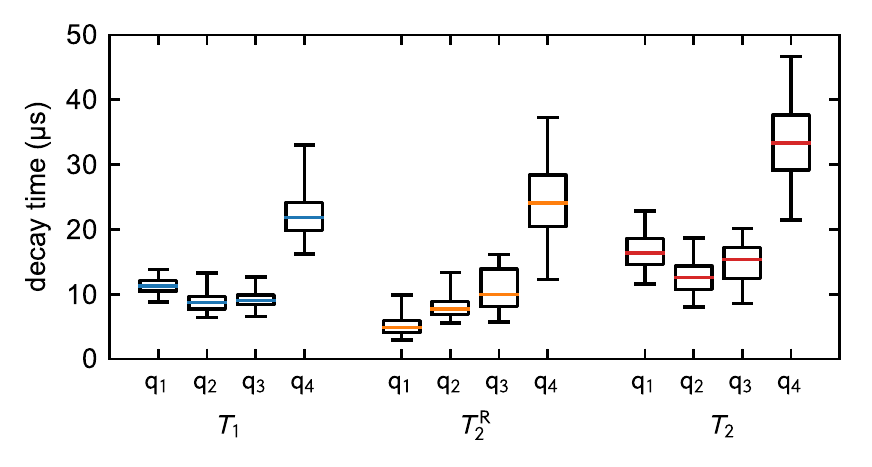}
	\caption{Long-term measurement of the qubits' decay times. The box encloses the second and third quartile, whereas the whiskers indicate $2\sigma$, i.e., 95\% of the data. The colored line indicates the median of each data set.}
	\label{fig:boxplot_char_times}
\end{figure}

Superconducting qubits are one of the most promising platforms to realize a universal quantum computer. In contrast to other popular qubit implementations, such as trapped ions \cite{Cirac1995}, cold atoms \cite{Bloch2008}, and NV centers \cite{Jelezko2006}, the properties of superconducting qubits are defined by a micro-fabricated electrical circuit. Consequently, most qubit parameters are adjustable by the circuit design and fabrication, and even the physical encoding of a quantum state is flexible \cite{Nakamura1999,Martinis2002,Chiorescu2003,Koch2007,Manucharyan2009}.
Superconducting qubits feature good coherence times in the range of $10-300\,\si{\micro\second}$ \cite{Paik2011,Barends2013,Nersisyan2019,Place2020}, which is long enough for several hundred to thousand qubit gates \cite{Kjaergaard2020}. Most recently, quantum advantage was for the first time demonstrated on a processor consisting of superconducting transmon qubits with an average lifetime of $T_{1} = 16\,\si{\micro\second}$\cite{Arute2019}.             
\\
The centerpiece of most superconducting qubits are Josephson junctions (JJ) serving as nonlinear elements. Their nonlinearity allows for the isolation of two of the circuit's quantum levels, usually ground, and first excited state, which may then be used as logical quantum states for computation.
Currently, several different techniques are employed to generate the superconductor-insulator-superconductor interface of the JJ. Most processes rely on electron-beam lithography as smaller areas enable lower loss in the JJ \cite{Martinis2005, Weides2011a}. In the commonly used shadow-evaporation processes, free standing bridges \cite{Dolan1977} or overhangs \cite{Lecocq2011}, and multi-angle evaporation are exploited to generate the desired interface in situ. One drawback of these techniques is a systematic angle dependent parameter spread across larger wafers, where great efforts are necessary to mitigate this spread \cite{Foroozani2019, Kreikebaum2019}. The need for point-like evaporation sources limits the applicable materials and growth processes. When polymer masks are employed in favor of hard masks \cite{Dolata2003,Tsioutsios2020} the superconductor choice is further restricted to metals with low melting temperatures. Additionally, the JJ can suffer from an outgasing of the resist.
\\
An alternative to shadow-mask technology are overlap JJ, which do not rely on angle dependent evaporation, and therefore promise superior scalability. Early implementations of micron sized overlap JJ with superconducting qubits suffered significantly from dielectric loss \cite{Steffen2006, Weides2011, Braumuller2015}. More recently, qubits with nanoscaled contacts feature coherence properties competitive with those stemming from shadow-evaporation techniques \cite{Wu2017}.
\\
However, current fabrication processes still rely on double resist stacks, and lift-off steps, limiting processing yield and presenting a potential source of contamination during the deposition \cite{Pop2012, Quintana2014}.
\\
In this work, we implement a subtractive process for patterning the JJ, where both electrodes are structured using etching rather than lift-off, allowing for smaller, more coherent contacts. Eliminating the resist from the evaporation chamber opens the door to homogeneous deposition, the addition of reactive gases, and evaporation at elevated temperatures. Consequently, new electrode materials, or epitaxial growth can be explored \cite{Fritz2019}. We demonstrate our fabrication platform using Al-AlO$_{x}$-Al JJ.
\begin{figure*}
	\centering
	\includegraphics{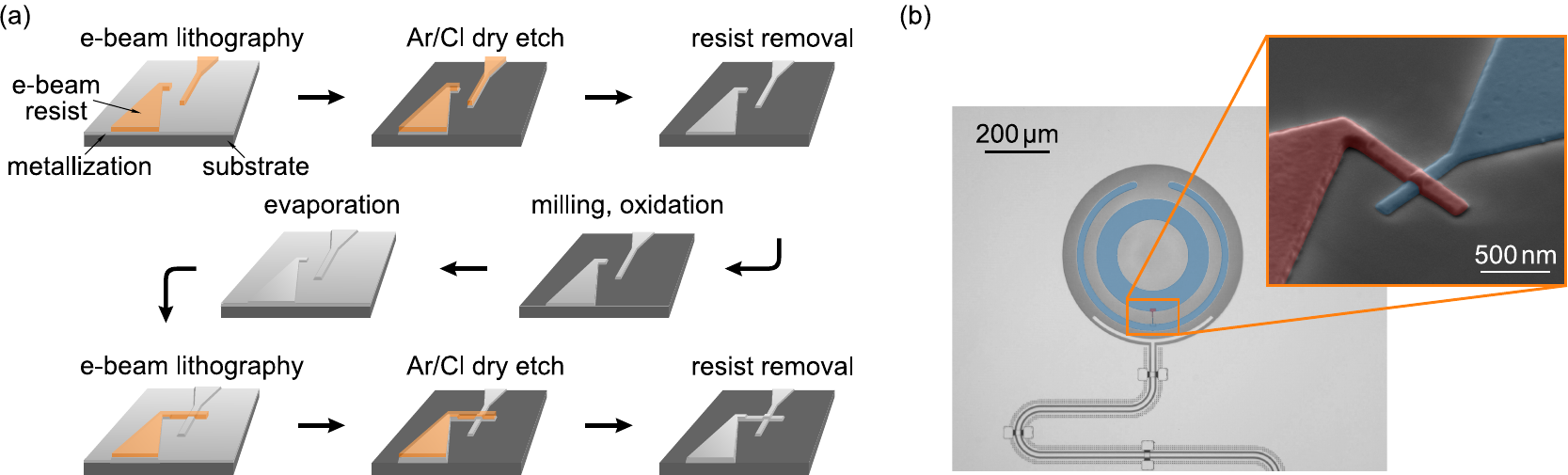}
	\caption{a) Sketch of the fabrication process. The process solely relies on two etching steps to pattern the electrodes of the JJ. The lithography for the top layer is performed identically to that of the bottom electode. b) False color image of the concentric transmon qubit, and SEM micrograph of the JJ. The bottom layer of aluminum is highlighted in blue. The top layer, forming the second JJ electode, is colored in red. The JJ electrodes have a width of $\sim\!\,180\si{\nano\meter}$.}
	\label{fig:fab-sketch}
\end{figure*}
Using an array of test contacts, we study aging of the room temperature resistance, which on average increases by only by 1.6\% over 6 months. Transmon qubits fabricated with this technique show good coherence properties, where the life-, and coherence times of our best device exceed on average \SI{20}{\micro\second}. The process is fully compatible with modern nanofabrication methods, making it an important ingredient for large scale fabrication of superconducting quantum processors.
\\
A schematic of the fabrication process is displayed in Fig.~\ref{fig:fab-sketch}a.
In the first step, a c-plane sapphire wafer is covered with aluminum at a thickness of \SI{50}{\nano\meter}, evaporated at a rate of $1\,{\rm nm/s}$. This layer defines the main structures of the circuit, as well as the bottom electrode of the JJ. Following, the latter is patterned using electron-beam lithography with $\sim 180\,\si{\nano\meter}$ thick PMMA resist. However, any resist with sufficient resistance to the etching plasma and a satisfactory resolution may be employed. A positive resist reduces electron-beam writing times. Subsequently, the structures are transferred to the aluminum film by reactive ion etching with an Ar/Cl plasma. The plasma is generated inductively using an rf-field with $100\,\si{\watt}$ at a gas flow of $15\,\text{cm}^3/\text{min}$ (sccm) argon and $2\,\text{sccm}$ chlorine gas, and is accelerated with a power of $100\,\si{\watt}$. After etching, the remaining resist is removed with a combination of ultrasonic cleaning, acetone, and N-ethyl-pyrrolidone. Milling, oxidation, and deposition of the top electrode are performed in situ, in a Plassys\textsuperscript{TM} MEB550S evaporation machine. First, resist residuals are incinerated in a 30 second Ar/O plasma. The native oxide on the aluminum film is removed by Ar sputtering for 180 seconds \cite{Grunhaupt2017a}. Immediately afterwards, the AlO$_x$ tunnel barrier is grown in a controlled manner by dynamic oxidation for 30 minutes, admitting a continuous flow of~$12\,\text{sccm}$~O$_2$ to the load lock, at chamber pressure of $p_{\rm LL}\approx 0.195\,\si{\milli\bar}$. The \SI{80}{\nano\meter} thick aluminum top layer is deposited in vacuum at a rate of $1\,{\rm nm/s}$. Analogously to the bottom electrode, the top layer is patterned with electron-beam lithography and an Ar/Cl plasma. Finally, larger structures can be applied using optical lithography. We note, that this process leaves us with a stray junction, which was shown to have a negative impact on qubit coherence times \cite{Lisenfeld2019}. This effect can be mitigated by employing a bandaging technique, which shorts the dielectric of the stray junction \cite{Dunsworth2017}.
\\
Using SEM imaging, we identify a process bias of $\sim 10$\% towards reduced junction edge width. Most likely, the chlorine introduces an isotropic component of the etching plasma, causing an under-etching and sloped side-walls of the aluminum films, thus reducing the width of the contact electrodes. In room temperature measurements, we find a normal state resistance times area product of ${R_{\rm n}A=(0.47\pm0.10)\,\si{\ohm\micro\meter}^2}$ across 36 test contacts with varied size, fabricated in the same batch as our qubits. After aging for $\sim 6$ months this value increased by about 1.6\%, see Fig.~\ref{fig:Rn-hist}. This indicates clean JJ interfaces \cite{Pop2012}. For additional information see supplementary material.
\begin{figure}[b]
	\includegraphics{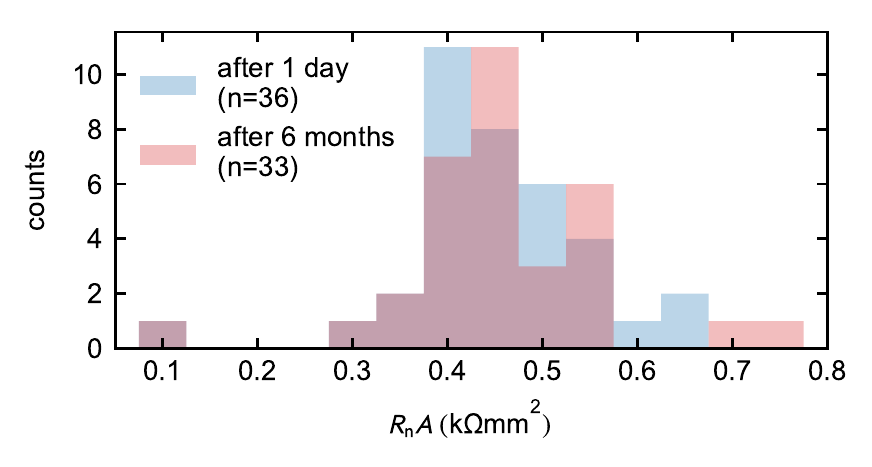}
	\caption{Room temperature resistance measurements of the test JJ, before, and after aging. The resistance is normalized to the JJ area. We vary the latter by increasing the JJ edge width from \SI{150}{\nano\meter} to \SI{300}{\nano\meter} in \SI{50}{\nano\meter} increments. The mean of $R_{\rm n}A$ increased by only 1.6\% after $\sim 6$ months of aging in ambient conditions.}
	\label{fig:Rn-hist}
\end{figure}
The spread in resistance is similar to that found in shadow evaporated junctions (before meticulous process optimization). It is likely to be caused by the nonuniformity of the electrode edges, constituting $\sim 25-40$\% of the total JJ area, due to an isotropic etching component caused by the chlorine. In the future, the spread in normal state resistance can be mitigated by reducing the thickness of both top and bottom electrode, and thereby the duration of the dry etch and effects of under-etching. This also enables the use of thinner electron-beam resists with better resolution. In combination, this allows to decrease overlap area, a crucial step for reducing dielectric loss in the JJ.
\\
Using the recipe described above, we fabricate a sample comprising two conventional (devices q$_1$ and q$_2$), and two concentric transmon qubits\cite{Braumuller2016} (devices q$_3$ and q$_4$), embedded in a coplanar microwave environment, see Fig.~\ref{fig:fab-sketch}b. A micrograph of the whole chip, our approach in identifying the qubits, and details on the qubit fabrication can be found in the supplementary material. For readout purposes, the qubits are capacitively coupled to a distributed $\lambda/4$-resonator, which are addressed in reflection measurements. The qubit population is determined by the dispersive shift of the respective readout resonator's frequency \cite{Blais2004,Wallraff2004}. Table \ref{table:device_parameters} summarizes the essential parameters of all four devices, which were extracted using spectroscopy measurements. The qubit-resonator coupling was calculated from the dispersive shift of the corresponding readout resonator.
\begin{ruledtabular}
	\begin{table}
		\caption{Device parameters in MHz. For each qubit, this includes the frequency of the readout resonator $\omega_{\rm r}$, and the first qubit transition $\omega_{01}$, as well as the qubit anharmonicity $\alpha$, the coupling strength $g$ to the readout resonator, and the resulting disperive shift $\chi$.}
		\begin{tabular}{c c c c c c}
			device &$\omega_{\rm r}/2\pi$ & ${\omega_{01}/2\pi}$ & ${\alpha/2\pi}$ & $\chi_{01}/2\pi$ & ${g/2\pi}$ \\ \hline
			q$_1$&6460 & 3548 & $-257$& 0.915 & $47.5$ \\
			q$_2$&6632 & 3950 & $-262$& 0.514 & $45.6$ \\
			q$_3$&6462 & 3161 & $-294$& 0.350 & $50.6$ \\
			q$_4$&6457 & 3324 & $-300$& 0.774 & $49.3$ \\
		\end{tabular}
		\label{table:device_parameters}
	\end{table}
\end{ruledtabular}
\\
We measure the lifetime $T_1$, Ramsey decay time $T_2^{\rm R}$, and spin-echo decay time $T_2$ of all qubits over several hours. By employing an interleaved measurement scheme, we resolve slow fluctuations of the qubit frequencies, life-, and coherence times \cite{Schlor2019,Burnett2019,Hong2020}. For each qubit, the combined measurement of a set of $T_1$, $T_2^{\rm R}$ and $T_2$ takes $\sim30\,\si{\second}$ for $10^3$ point averages. A typical measurement trace from device q$_4$ is displayed in Fig.~\ref{fig:exemplary_decay_times}.
Here, the $\pi/2$-pulse in the Ramsey-sequence was detuned by $\sim 50\,\si{\kilo\hertz}$, which results in characteristic oscillations in the laboratory frame of reference. For a detailed sketch of the measurement setup, see supplementary material.
\\
A comparison of the coherence properties of all four qubits is displayed in Fig.~\ref{fig:boxplot_char_times}, in a boxplot. For the full distribution we refer to the supplementary material. Data sets with a fit error exceeding 50\% are neglected. Furthermore, we exclude traces where either of the coherence times exceeds $2T_1$, or where $T_2^{\rm R} > T_2$. The median lifetime $\tilde{T}_1$, coherence time  $\tilde{T}_2^{\rm R}$,  and  $\tilde{T}_2$ of all four devices are summarized in table \ref{table:char_times}.
\begin{ruledtabular}
	\begin{table}
		\caption{Overview of the qubits' median lifetime $\tilde{T}_1$, and median coherence times $\tilde{T}_2^{\rm R}$, and $\tilde{T}_2$ in $\si{\micro\second}$.}
		\begin{tabular}{c r r r}
			device &$\tilde{T}_1$ &$\tilde{T}_2^{\rm R}$&$\tilde{T}_2$ \\ \hline
			q$_1$&11.3 & 4.9 & 16.3 \\
			q$_2$&8.7 & 7.7 & 12.5 \\
			q$_3$&9.0 & 9.9 & 15.3 \\
			q$_4$&21.8 & 24.1 & 33.3 \\
		\end{tabular}
		\label{table:char_times}
	\end{table}
\end{ruledtabular}
Devices q$_1$-q$_3$ perform slightly worse than q$_4$. A potential cause is aluminum residuals in the vicinity of the JJ. The performance of device q$_4$ is close to the results of qubits with JJ made from shadow evaporation or lifted overlaps, with state of the art transmon qubits performing better by a factor of $\sim 5-10$. Here, our devices would profit from a reduction of surface loss \cite{Gambetta2017}, and optimization of the electrode materials \cite{Place2020}.

In conclusion, we established a technique for the subtractive fabrication of coherent JJ. Our recipe does generally not rely on lift-off processes, is angle independent, and tolerates depositions at elevated temperatures and in reactive gases. Furthermore, our approach is extremely flexible with respect to the electrode materials, and growth processes. The junctions feature low aging of the normal state resistance, indicating clean JJ interfaces. These are important ingredients for streamlined and large scale processing platform of superconducting quantum processors. We demonstrated good coherence properties of four transmon qubits with subtractive JJ, where the average life- and coherence times of our best device exceed $20\,\si{\micro\second}$.\\

\begin{acknowledgments}
	Cleanroom facilities use was supported by the KIT Nanostructure Service Laboratory (NSL). We thank A. Lukashenko for technical support. This  work  was  supported  by  the  European Research  Council  (ERC)  under  the  Grant  Agreement No.  648011, Deutsche Forschungsgemeinschaft (DFG) projects INST 121384/138-1FUGG and WE 4359-7, EPSRC Hub in Quantum Computing and Simulation EP/T001062/1, and the Initiative and Networking Fund of the Helmholtz Association. AS acknowledges support from the Landesgraduiertenf\"orderung Baden-W\"urttemberg (LGF), JDB acknowledges support from the Studienstiftung des Deutschen Volkes. TW acknowledges support from the Helmholtz International Research School for Teratronics (HIRST). AVU acknowledges partial support from the Ministry of Education and Science of the Russian Federation in the framework of the Increase Competitiveness Program of the National University of Science and Technology MISIS (contract No. K2-2020-017).
\end{acknowledgments}

\section*{Availability}
This article may be downloaded for personal use only. Any other use requires prior permission of the author and AIP Publishing. This article appeared in \cite{Stehli2020} and may be found at \href{https://aip.scitation.org/doi/10.1063/5.0023533}{https://aip.scitation.org/doi/10.1063/5.0023533}. The data that support the findings of this study are available from the corresponding author upon reasonable request.

\section*{References}
\bibliography{Stehli_Coherent_qubits_from_subtractive_JJ_process}
\clearpage

\begin{widetext}
\section*{Supplementary material}
\renewcommand{\thepage}{S\arabic{page}}
\renewcommand{\thesection}{S\arabic{section}}
\renewcommand{\thetable}{S\arabic{table}}
\renewcommand{\thefigure}{S\arabic{figure}}
\setcounter{figure}{0}
\setcounter{page}{1}
\subsection*{Qubit life-, and coherence time distribution}
The distributions of all qubits' life-, and coherence times are displayed in Fig.~\ref{fig:char_time_hists}. As described in the main text, data sets with a fitting error exceeding 50\% are excluded. Finally, we also exclude unphysical data sets, i.e., where $T_2 > 2 T_1$, $T_2^{\rm R} > 2 T_1$, or $T_2^{\rm R} > T_2$, which may occur due to fit errors or qubit fluctuations during the measurement. For each qubit table \ref{table:measurement_times} summarizes the number $N_{\rm tot}$ of total, and successful measurements $N$ (converged fit), as well as the duration of the measurements.
\begin{figure}[h]
	\includegraphics{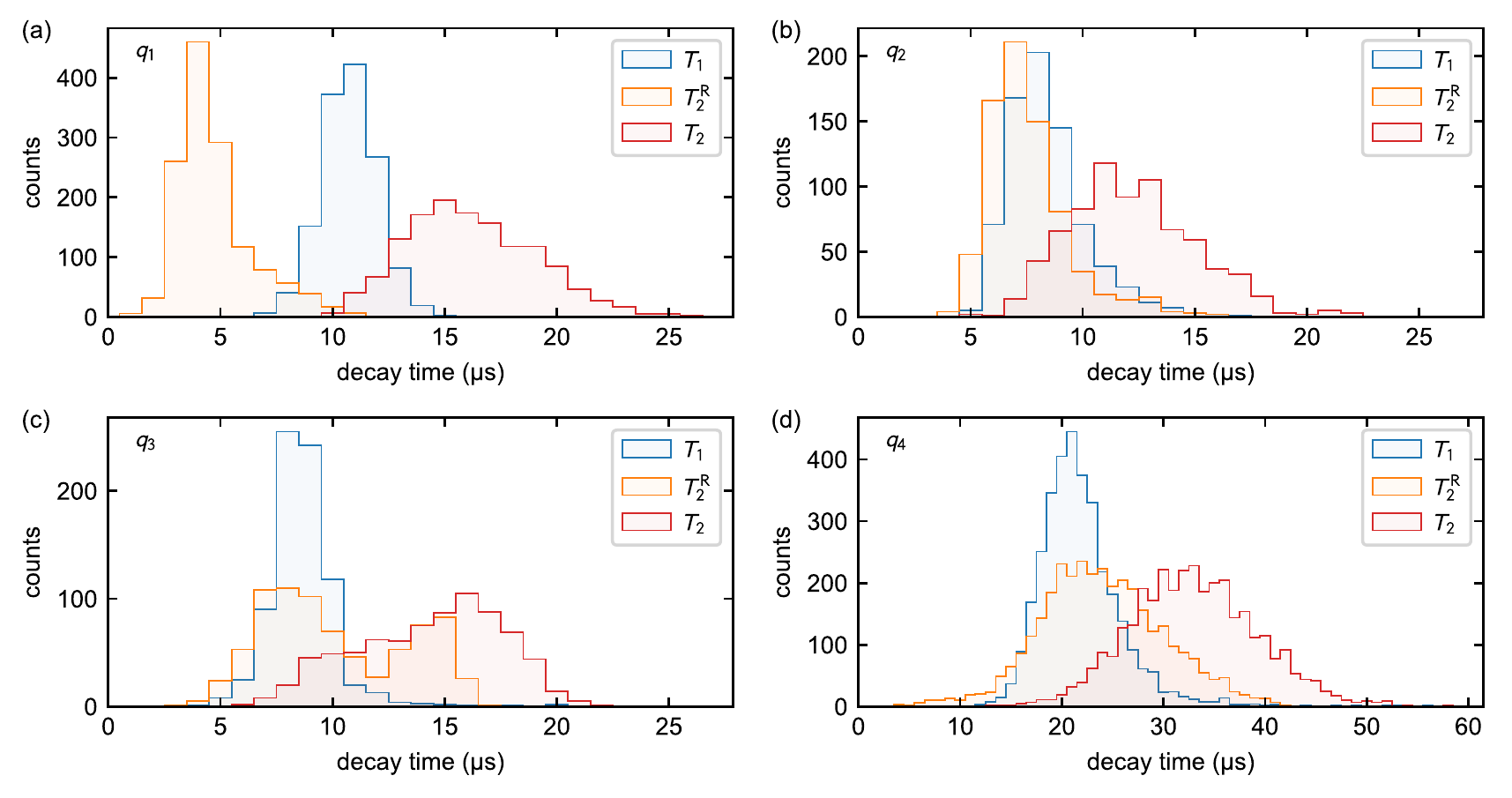}
	\caption{Life-, and coherence time distribution for all qubits (q$_1$-q$_4$ from a-d).}
	\label{fig:char_time_hists}
\end{figure}
\begin{table}[h]
	\centering
	\caption{Overview of the measurement duration, total number of measurements $N_{\rm tot}$, and successful number of measurements $N$.}
	\begin{tabular}{c c r c}
		\hline \hline
		device & measurement duration (h) & \hspace{15pt}$N_{\rm tot}$\hspace{5pt} & \hspace{15pt}$N$\hspace{15pt} \\ \hline
		q$_1$& 29.8 & 1700 & 1589 \\
		q$_2$& 17.6 & 2000 & 1382 \\
		q$_3$& 35.3 & 2502 & 1744\\
		q$_4$& 69.6 & 10337 & 7013 \\ \hline \hline
	\end{tabular}
	\label{table:measurement_times}
\end{table}

\subsection*{Microwave setup}
In Fig.~\ref{fig:setup}a the microwave setup for spectroscopy and time domain experiments, as well as the wiring inside the cryostat, are summarized. In our time domain experiments, we employ single-sideband mixing to generate the desired pulses for qubit manipulation and readout. We use the same local oscillator for up and down conversion of the readout signal. The signal is digitized at a rate of 500 MS/s by an ADC card installed in our measurement PC.
\\
As is common practice, we use a combination of attenuators on various temperature stages, high-pass filter, infrared filters and circulators/isolators to protect the sample form external radiation.
\\
Figure \ref{fig:setup}b gives an overview of the on-chip microwave setup. Each qubit is coupled capacitively to a dedicated $\lambda/4$-readout resonator. The resonators couple inductively to a transmission line, which connects to the output of the sample holder on one end, and is terminated to ground on the other.
\begin{figure}[h]
	\includegraphics{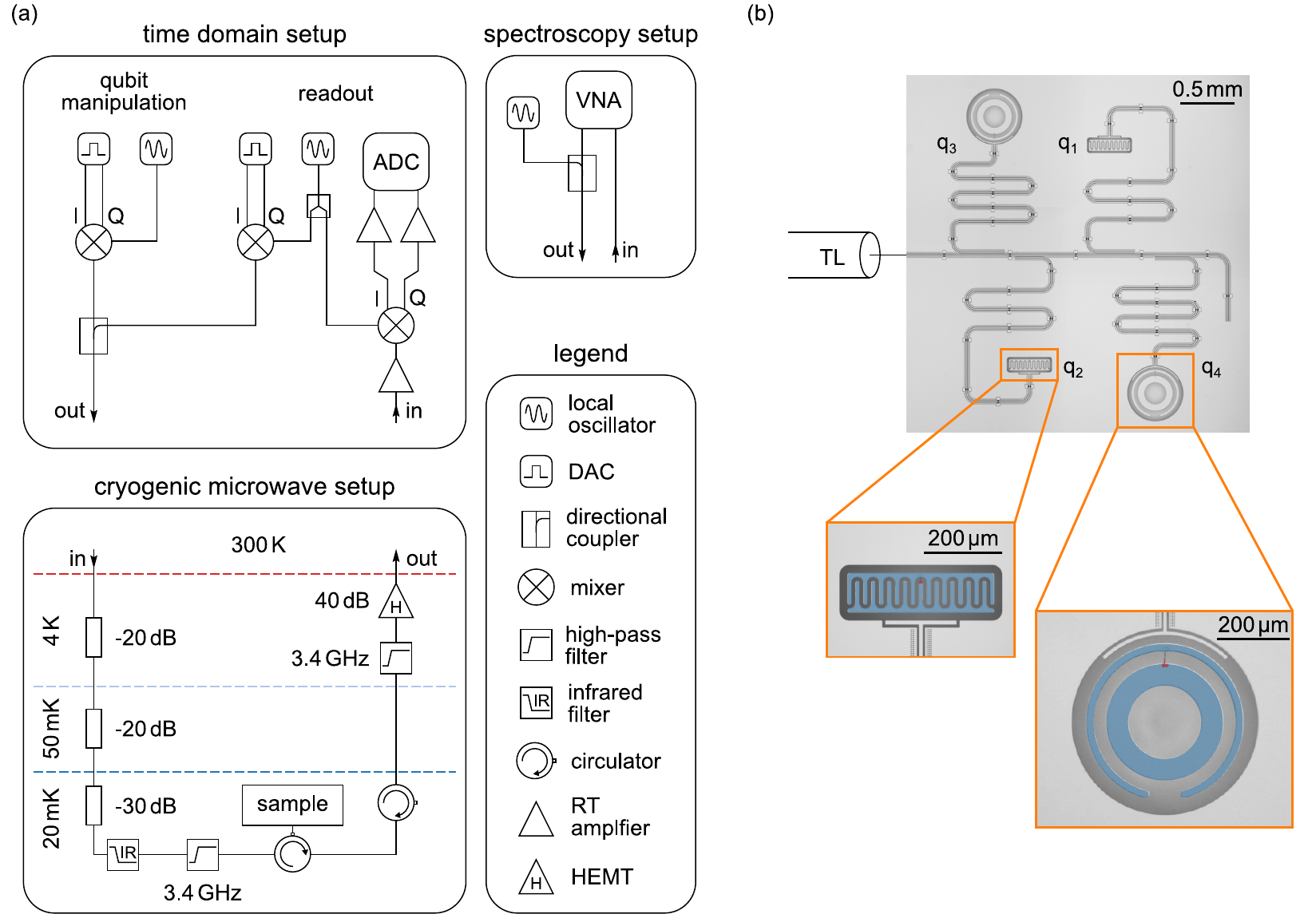}
	\caption{a) Schematic of the time domain and spectroscopy setup, as well as the cryogenic microwave setup. I and Q are in-phase and quadrature components of the microwave signal. b) Overview of the sample. The qubit electrodes in the zoom in are highlighted in blue. The transmission line (TL) connects to the sample box output on the left-hand side, and terminates to ground on the right-hand side.}
	\label{fig:setup}
\end{figure}

\subsection*{Normal state JJ resistance and aging}
In room temperature measurements we characterize a set of 36 JJ fabricated in the same batch as the qubits. We measure the normal state resistance $R_{\rm n}$ in a 2-point measurement. Compared to a 4-point probe method the systematic error acquired is negligible due to the large resistance ($> 4\,\si{\kilo\ohm}$) of the contacts. For better comparability, we multiply the normal state resistance with the area of the overlap. Immediately after the fabrication, our measurements yield $R_{\rm n}A = (0.474 \pm 0.099)\,\si{\kilo\ohm\milli\meter}^2$. In order to quantify JJ aging, we repeat the characterization after $\sim 6$ months. During this time, the samples were stored under ambient conditions. Neglecting measurement inaccuracies, we find a slight increase of about 1.6\% to $R_{\rm n}A = (0.482 \pm 0.108)\,\si{\kilo\ohm\milli\meter}^2$.

\subsection*{Qubit fabrication}
In our qubit fabrication, the top electrode is deposited only in a small window above the JJ. Here, an optical lift-off process with S1805 resist was employed. We want to emphasize, that this lift-off step is not process relevant, and has the sole purpose of reducing the electron-beam exposure time. Using a negative tone resist, as done for some of our latest samples, renders it obsolete. The main structures of the qubit sample are also patterned optically, using S1805 resist and an Ar/Cl plasma. We admit an additional flow of~$1\,\text{sccm}$~of oxygen, which improves the edge roughness of our main structures. During the last process step, the JJ is protected by photo resist.

\subsection*{Resonator identification}
Due to the close proximity of the readout resonators in frequency space, an identification of the qubits proves challenging.  On a dummy chip with the same structures, we applied a drop of varnish on a readout resonator, shifting its transition frequency. Consequently, we can determine which qubit corresponds to the coated resonator. By repeating this process twice we identified two resonator-qubit pairs, see Fig.~\ref{fig:resonator_id}. The identity of the remaining qubits was inferred from their anharmonicity.
\begin{figure}[h]
	\includegraphics{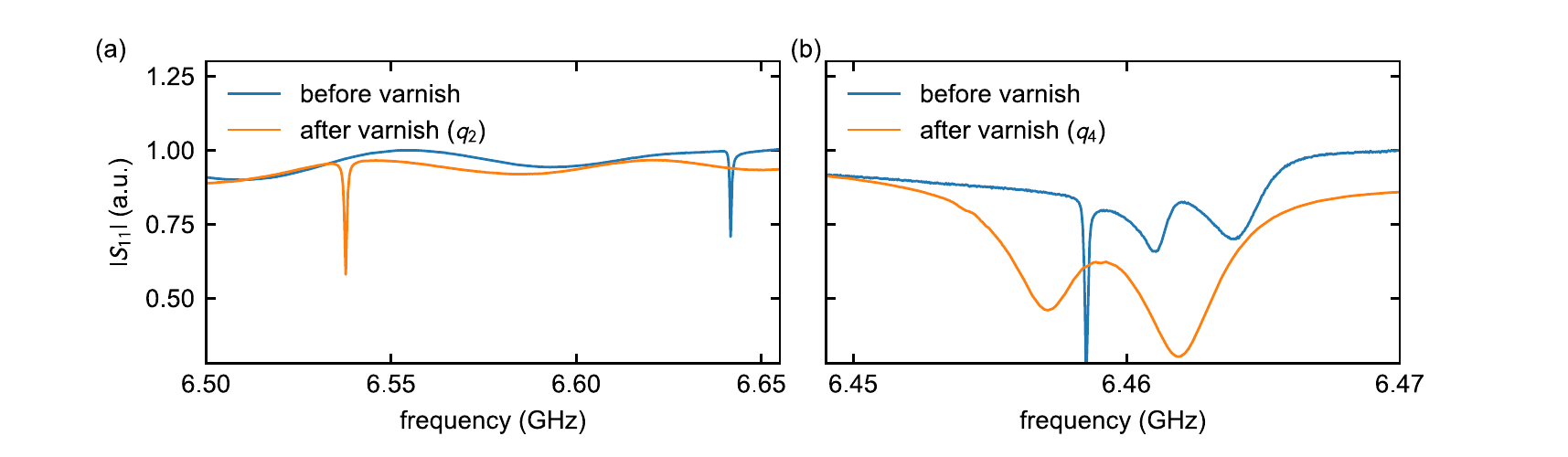}
	\caption{Varnish induced shift of the readout resonator frequencies. In order to identify the the qubits, we shift the frequency of one readout resonator by applying a drop of varnish and identify the change in the reflection coefficient $\left|S_{11}\right|$. a) Shift of the highest frequency resonator (conventional geometry). b) Subsequently, the lowest frequency resonator is shifted, revealing that the corresponding qubit is a concentric transmon.}
	\label{fig:resonator_id}
\end{figure}
\clearpage

\end{widetext}
\end{document}